\documentclass[]{spie}

\usepackage{graphics}

\begin{document}

\title{Optical vortex trajectories in an astigmatic and elliptical Gaussian beam} 

\author{Filippus S. Roux \\
Department of Electrical, Electronic and Computer Engineering, \\
University of Pretoria, Pretoria 0001, South Africa }


\maketitle

\begin{abstract}
An optical vortex, produced at one point in an optical beam, would propagate through an optical system to another point where the vortex can be used for some purpose. However, asymmetrical optical elements in such a system can cause astigmatism or at least distroy the rotational symmetry of the beam, which may affect the propagation of the vortex in an undesirable way. While an optical vortex in a rotationally symmetric, stigmatic Gaussian beam retains its initial morphology for as far as it propagates, the morphology of an optical vortex in an asymmetric or astigmatic Gaussian beam changes. The vortex can even be replaced by another with the opposite topological charge. We consider the behavior of single noncanonical vortices propagating in Gaussian beams that are asymmetric and/or astigmatic. General expressions for the vortex trajectories are provided. The locations of the flip planes and the evolution of the anisotropy of the vortex are considered for different non-ideal situations.
\end{abstract}



\section{INTRODUCTION}
\label{intro}

One of the significant advances that is being made in nano-technology is the ability to manipulate small objects and particles, even to the level of single atoms. A highly focused optical beam can be used as an `optical tweezer' with which single atoms can be moved around. On an even more sophisticated level, optical beams can also be used to rotate or twist small particles. In the early 90's Allen and coworkers discovered that an optical beam can carry {\it orbital angular momentum}\cite{oam} (in addition to spin angular mometum, which is associated with circular polarization). This angular momentum can be transferred to a small particle that absorbs the light.

Optical orbital angular momentum is associated with phase singularities, called {\it optical vortices}, which were discovered by Nye and Berry in the 60's.\cite{nb} An optical vortex is a point on the cross section of a beam (or a curved line moving along the beam). At this point the intensity of the beam is zero. Around this point the phase increases by an integer multiple of $2\pi$. This integer is called the {\it topological charge\/} of the vortex. Most of the time vortices only appear with topological charges of $\pm 1$. The two signs represent two opposite helicities, which indicate whether the phase increases in a clockwise or counterclockwise direction around the phase singularity.

The rate of increase of the phase around a vortex is not always a constant function of the azimuthal angle. For a {\it canonical\/} vortex the rate of increase is constant and the vortex has an isotropic (rotationally symmetric) appearance. One can regard a {\it noncanonical\/} vortex as the anisotropically scaled (or stretched) version of a canonical vortex. Together with the direction or orientation of the scaling, the amount of anisotropy represents the {\it morphology\/} of the optical vortex.

One can create an optical vortex in a beam with a {\it computer generated hologram}.\cite{cgh} This vortex then propagates along the beam to the point where it can be applied to rotate some small objects.\cite{grier} Unfortunately, the vortex does not always remain the same as it propagates. If there are more than one vortex in the beam they interact with each other and can even annihilate each other.\cite{koppel} Moreover, in the late 90's a group in spain\cite{cyl} performed an interesting experiment in which they demonstrated that the topological charge of an optical vortex can be inverted after the vortex bearing beam passed through a cylindrical lens. Hence, under certain circumstances, even if there is only one vortex in the beam, it can change its shape (morphology) and even its topological charge. So the topological charge in a beam is not always conserved during propagation. What happens is that two `vortices' are brought in from infinity by means of an edge phase dislocation to cancel the vortex in the beam and replace it with a vortex of opposite topological charge. This happens because of the astigmatism that the beam picked up from the cylindrical lens. The topological charge is still {\it locally} conserved in that the net inward (or outward) flow of topological charge through a closed surface is always zero. 

One of the consequences of this observation is that there is a set of special planes in such an astigmatic beam where the topological charge is inverted.\cite{soskin} One of the questions one can ask is where these planes are located and what these locations depend on. Here we report on the results of this investigation. It is well known that a single noncanonical vortex that propagates in a rotationally symmetric, stigmatic Gaussian beam retains its morphology for as far as it propagates.\cite{skew} When the rotational symmetry is broken or the beam picked up some astigmatism this situation is no longer true -- the morphology of the vortex evolves during propagation. If the Gaussian beam is asymmetrical (elliptical) but not astigmatic and the vortex is canonical in the waist of the Gaussian beam, then, although the vortex becomes noncanonical, it does not change its topological charge. If it is noncanonical in the waist, there can be a region over which the topological charge of the vortex is inverted. A more detail analysis of these conditions is provided below. If the beam is astigmatic but symmetric in the sense that the Rayleigh distances associated with the two transverse directions are equal, then the vortex can invert its topological charge irrespective of its initial morphology. Hence it is both the morphology of the vortex and the astigmatism of the beam that determines whether topological charge inversion occurs.

The rest of the paper is organized as follows. We define our notation for the beam and the morphology of the vortex in Section \ref{notas}. The general expressions for the trajectory of a single vortex with an arbitrary morphology propagating in a general asymmetrical and/or astigmatic Gaussian beam are provided in Section \ref{traj}. The zeros of the denominator of these trajectories provide the locations for the flip planes. In Section \ref{flipp} these flip plane locations are investigated for various situations. We provide expressions for the anisotropy of the vortex as a function of the propagation distance in Section \ref{anis} to show that the anisotropy vanishes and changes sign at the locations of the flip planes, which confirms that topological charges inversion occurs at the flip plane locations. Conclusions are provided in Section \ref{concl}.

\section{NOTATION}
\label{notas}

A vortex bearing Gaussian beam can be expressed as a product of a complex valued polynomial function (i.e.\ the {\it prefactor}) and a Gaussian function. We'll define the propagation direction as the $z$-direction and the transverse directions as the $x$- and $y$-directions. Here, we'll use normalized coordinates $u=x/\omega_0$, $v=y/\omega_0$ and $t=z/\rho$, where $\omega_0$ and $\rho$ represent the waist size and the associated Rayleigh range of the Gaussian beam, respectively.

An astigmatic Gaussian beam has two different waists (foci) for two orthogonal transverse directions, which we define as the $x$- and $y$-directions, respectively. We define a {\it central plane\/} as the plane that is exactly halfway between the two focal planes of the astigmatic beam. For a stigmatic beam the central plane coincides with the waist of the Gaussian beam. One can then quantify the amount of astigmatism by a parameter $\tau$, which is defined as the distance from the central plane to one of the focal planes, normalized with respect to the nominal Rayleigh range (defined below). Note that $\tau$ is a dimensionless parameters that is given as the ratio of the two most significant scale parameters in the beam.

A Gaussian beam can be asymmetric in the sense that the size of the beam waist may be different for the $x$- and $y$-directions, respectively. In this case we define, for the purpose of normalization, a {\it nominal\/} waist size $\omega_0 = \sqrt{\omega_x \omega_y}$, where $\omega_x$ and $\omega_y$ are the waist sizes in the $x$- and $y$-directions, respectively. The associated nominal Rayleigh range is then $\rho=\sqrt{\rho_x \rho_y}$. We quantify the asymmetry of the beam by a parameter $\gamma = \omega_x/\omega_y = \sqrt{\rho_x/\rho_y}$. For definiteness we'll assume that $\gamma \geq 1 \Rightarrow  \omega_x \geq \omega_y$. For $\gamma=1$ the beam is called {\it symmetric\/} (which is not to be confused with {\it rotationally\/} symmetric beams where we also have $\tau=0$).

The prefactor for a Gaussian beam with a single on-axis noncanonical vortex is given by \cite{aniso}
\begin{equation}
V(u,v) = {1\over\sqrt{2}} [\xi (u + i v) + \zeta (u - i v)] ,
\label{vort}
\end{equation}
where the {\it morphology} of the vortex is given in terms of\cite{ss}
\begin{eqnarray}
\xi & = & \cos(\psi/2) \exp(-i\phi/2) \\
\zeta & = & \sin(\psi/2) \exp(i\phi/2) ,
\label{morfhoeke}
\end{eqnarray}
with $\psi$ and $\phi$ being the {\it morphology angles}, representing the anisotropy and orientation, respectively. The sign of $\cos\psi$ represents the {\it topological charge\/} of the vortex. For an off-axis vortex the expression in Eq.~(\ref{vort}) can be shifted to an arbitrary location $(u_0,v_0)$.

Hence, in its central plane a general asymmetric astigmatic Gaussian beam with a single off-axis noncanonical vortex can be expressed by
\begin{equation}
g(x,y) = {1\over\sqrt{2}} \left[ (\xi+\zeta)(u-u_0) + i (\xi-\zeta)(v-v_0) \right] \exp \left(-{u^2\over\gamma+i\tau}-{v^2\over 1/\gamma-i\tau}\right) .
\label{gauss}
\end{equation}
The expression at other points along the direction of propagation can be obtained with the Fresnel transform\cite{goodman} of this expression.

\section{TRAJECTORIES}
\label{traj}

The prefactor changes as a function of $t$. The zeros of the prefactor give the locations of the vortex as a function of the propagation distance -- the vortex trajectories. For a general asymmetric astigmatic Gaussian beam with a single off-axis noncanonical vortex, the vortex trajectory is given by
\begin{equation}
u(t) = {\left[ \gamma^2 + (\tau-t)^2 \right] \left \{ \gamma \left[ (C\tau\gamma-A) u_0 + (1-B) v_0 \right] t + C (\tau^2\gamma^2+1) u_0 \right \} \over \left[ A \tau \gamma (1+\gamma^2) + C (1-\tau^2) \gamma^2 \right]t^2 - \left[ A \gamma (\gamma^2-1) (1-\tau^2) - C \tau (\gamma^4-1) \right] t + C (\tau^2\gamma^2+1) (\tau^2+\gamma^2)} ,
\label{traju}
\end{equation}
\begin{equation}
v(t) = {-\left[ 1 + (t+\tau)^2\gamma^2 \right] \left \{ \left[ (1+B) \gamma u_0 + (C\tau-A\gamma) v_0 \right] t - C (\tau^2+\gamma^2) v_0 \right \} \over \left[ A \tau \gamma (1+\gamma^2) + C (1-\tau^2) \gamma^2 \right]t^2 - \left[ A \gamma (\gamma^2-1) (1-\tau^2) - C \tau (\gamma^4-1) \right] t + C (\tau^2\gamma^2+1) (\tau^2+\gamma^2) }
\label{trajv}
\end{equation}
where
\begin{equation}
A = \sin\psi_0 \sin\phi_0 \label{akon} ~~~~~ B = \sin\psi_0 \cos\phi_0 \label{bkon} ~~~~~
C = \cos\psi_0 \label{ckon} ,
\label{abc}
\end{equation}
with $\psi_0$ and $\phi_0$ being the morphology angles in the central plane.

\section{FLIP PLANE LOCATIONS}
\label{flipp}

The values of $t$ for which the denominator of the trajectory functions, given in Eqs.~(\ref{traju}--\ref{abc}), become zero represent points along the propagation direction where the vortices move to transverse infinity. As we'll show below the topological charges are inverted in the planes at these locations. These planes are therefore called {\it flip planes}. Since the denominator is quadratic in $t$ there are in general two such planes in a beam. The existence of these planes in astigmatic Gaussian beams is in itself an interesing fact, because they do not in general coincide with any of the other special planes in the beam, such as the focal planes. 

Note that the denominator in Eqs.~(\ref{traju}) and (\ref{trajv}) is independent of the coordinates of the vortex in the central plane, $(u_0,v_0)$. It only depends on the astigmatism parameter $\tau$, the symmetry parameter $\gamma$ and the initial morphology of the vortex.

We now consider the locations of the flip planes under two special circumstances: for elliptical stigmatic beams ($\tau=0$) and for symmetric astigmatic beams ($\gamma=1$).

\subsection{Elliptical stigmatic beams}

First we consider the location of the flip planes in an elliptical stigmatic beam and thus set $\tau=0$ in the denominator of the trajectory functions. The locations of the flip planes are then given by
\begin{equation}
t_{\rm fp} = {T (\gamma^2-1) \pm \sqrt{T^2 \left( \gamma^2-1 \right)^2-4\gamma^2} \over 2\gamma} = K \pm \sqrt{K^2-1} ,
\label{flipdis}
\end{equation}
where
\begin{equation}
K = {T (\gamma^2-1) \over 2\gamma} = {T\over 2} \left( {\rho_x\over\rho} - {\rho_y\over\rho} \right)
\label{kparm}
\end{equation}
and
\begin{equation}
T = \tan\psi_0 \sin\phi_0 .
\label{tterm}
\end{equation}

\begin{figure}[ht]
\centerline{\scalebox{0.5}{\includegraphics{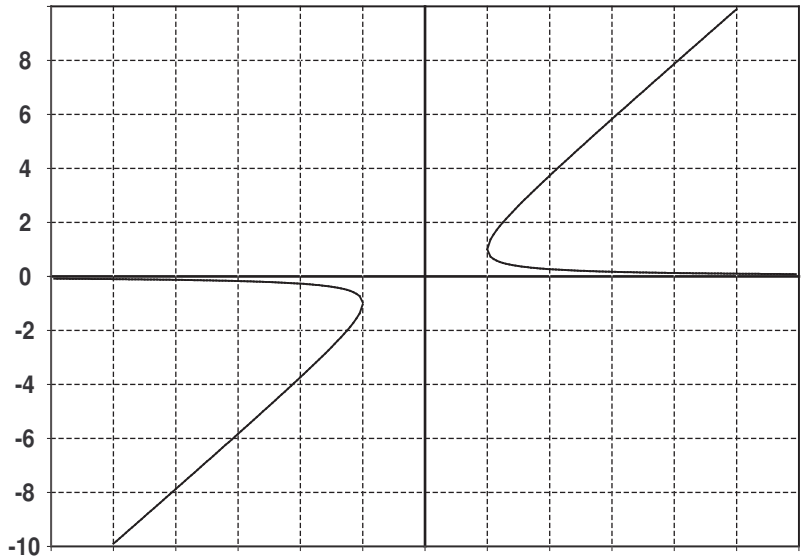}}}
\caption{The locations of the flip planes in a stigmatic beam as a function of $K$.}
\label{flipafst}
\end{figure}

The locations of the flip planes as a function of $K$ are shown in Fig.~\ref{flipafst}. One can see that the flip planes are always located on the same side of the central plane. In the extreme case ($K\rightarrow\infty$) when the vortex is an edge phase dislocation there is only one flip plane located in the central plane.

We see that there are only flip planes when 
\begin{equation}
|\tan\psi_0 \sin\phi_0| \geq {2\gamma \over \gamma^2-1} ,
\label{fpreq}
\end{equation}
which implies that a certain amount of anisotropy is required for the vortex in the initial plane to produce flip planes. A canonical vortex in the waist cannot undergo topological charge inversion in a stigmatic beam. Moreover, if the beam is symmetric ($\gamma=1$) in addition to being stigmatic ($\tau=0$) -- in other words, the beam is rotationally symmetric -- then a finite amount of anisotropy cannot satisfy the requirement of Eq.~(\ref{fpreq}).

\subsection{Symmetric astigmatic beams}

Now we consider the location of the flip planes in a symmetric astigmatic beam for which we set $\gamma=1$ in the denominator of the trajectory functions. In this case the locations of the flip planes are given by
\begin{equation}
t_{\rm fp} = \pm {1+\tau^2 \over \sqrt{\tau^2-2 T\tau-1}} ,
\end{equation}
with $T$ given in Eq.~(\ref{tterm}). We see that the flip planes only exist when 
\begin{equation}
\tau > T+\sqrt{T^2+1} .
\end{equation}
Note that the two flip planes are now located at equal distances from the central plane and they lie on opposite sides of the central plane.

The minimum flip plane distance for each value of $\tau$ occurs when
\begin{equation}
T = \tan\psi_0 \sin\phi_0 = {\tau(\tau^2-3)\over 3\tau^2-1} ,
\end{equation}
and this minimum distance is given by
\begin{equation}
t_{\rm fp} = \sqrt{3\tau^2-1} .
\end{equation}
This implies that the flip plane moves all the way to the central plane ($t=0$) when $\tau=1/\sqrt{3}$, for which $T\rightarrow\infty$. In other words, the vortex in the central plane becomes an edge dislocation.

\begin{figure}[ht]
\centerline{\scalebox{0.5}{\includegraphics{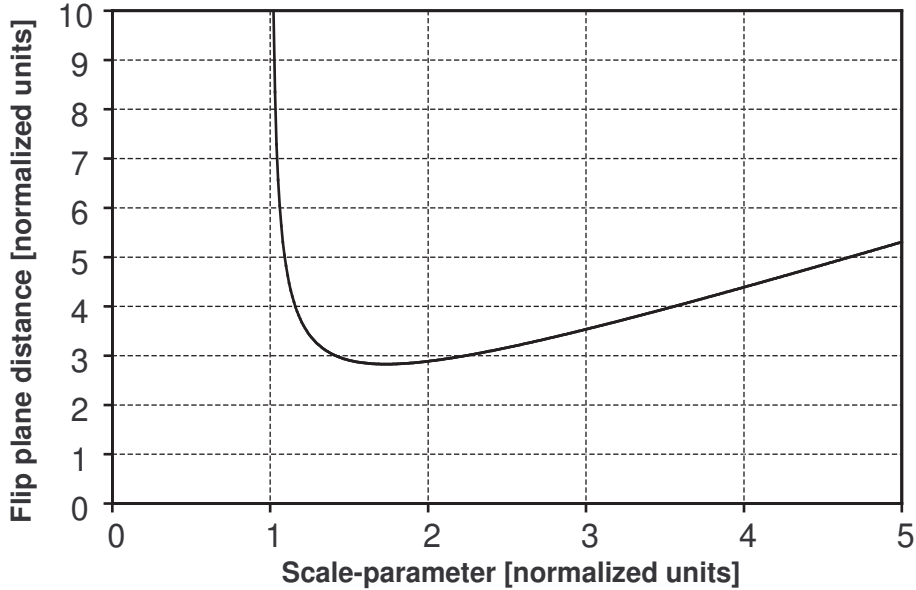}}}
\caption{The flip plane distance in a symmetric beam ($\gamma=1$) with a canonical vortex in the central plane ($T=0$), as a function of $\tau$.}
\label{flip}
\end{figure}

The flip plane distance is shown in Fig.~\ref{flip} as a function of $\tau$ for the canonical case ($T=0$). Here flip planes exist when $\tau\geq 1$. The flip plane distance starts at infinity for $\tau=1$, reaches a minimum at $\tau=\sqrt{3}$ and rises again to approach infinity for $\tau\rightarrow\infty$. The minimum flip plane distance at $\tau=\sqrt{3}$ is $t=2\sqrt{2}$. 

\section{ANISOTROPY}
\label{anis}

The significance of the existence of the flip planes lies in the fact that the vortices change their topological charge in those planes. One can show this by computing the anisotropy of the vortex as a function of the propagation distance. The anisotropy of a vortex in a complex valued function $f$ is given by\cite{koppel}
\begin{equation}
\cos\psi = {|\partial_- f|^2-|\partial_+ f|^2 \over |\partial_- f|^2+|\partial_+ f|^2} ,
\end{equation}
evaluated at the location of the vortex, where
\begin{equation}
\partial_{\pm} = \frac{1}{\sqrt{2}} \left( {\partial\over\partial x} \pm i {\partial\over\partial y} \right) .
\end{equation}

For an asymmetric astigmatic Gaussian beam with a vortex of arbitrary morphology in the central plane, the anisotropy of the vortex as a function of the propagation distance is given by
{\small
\begin{equation}
\cos\psi(t) = {2\left \{ \left[ A \tau \gamma (1+\gamma^2) + C (1-\tau^2) \gamma^2 \right]t^2 - \left[ A \gamma (\gamma^2-1) (1-\tau^2) - C \tau (\gamma^4-1) \right] t + C (\tau^2\gamma^2+1) (\tau^2+\gamma^2) \right \} \over \left[ B (\gamma^4-1) + \gamma^4 + 2 \gamma^2 \tau^2 + 1 \right] t^2 + 2 \tau \left[ B (\gamma^4+2\gamma^2\tau^2+1)+\gamma^4-1 \right] t + 2 (1+\gamma^2\tau^2) (\gamma^2+\tau^2)} .
\label{anif}
\end{equation}}
In the central plane the anisotropy is given by $\cos\psi_0$. From the fact that the numerator of the anisotropy function in Eq.~(\ref{anif}) is proportional to the denominator of the trajectories, provided in Eqs.~(\ref{traju}) and (\ref{trajv}), one can conclude that the anisotropy becomes zero. The topological charge is given by the sign of the anisotropy. One can therefore see that the topological charge changes sign at the flip planes.

Next the anisotropy is considered for elliptical stigmatic beams and for symmetric astigmatic beams.

\begin{figure}[ht]
\centerline{\scalebox{0.5}{\includegraphics{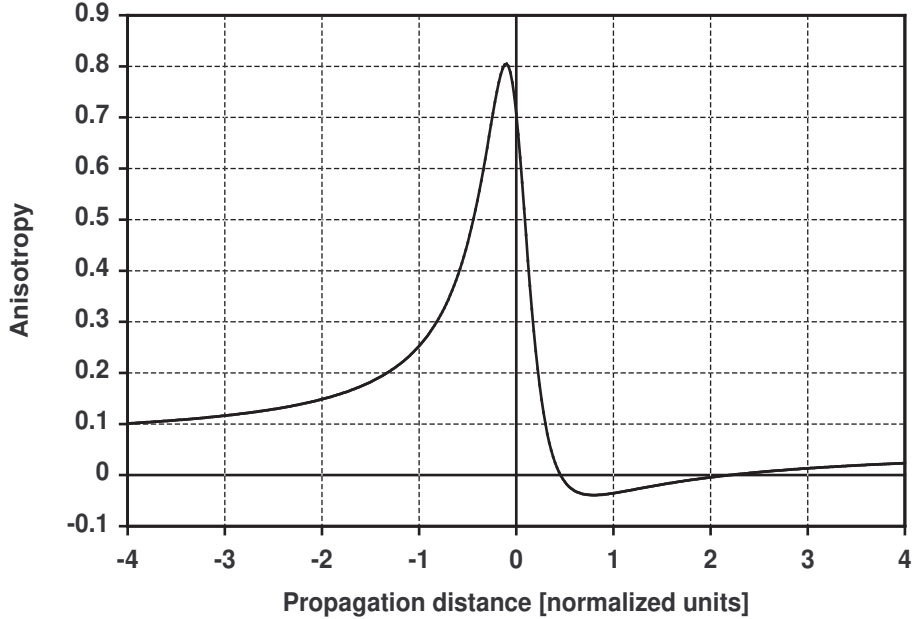}}}
\caption{The evolution of the anisotropy ($\cos\psi(t)$) for a stigmatic asymmetric beam with $\gamma=4$ and a vortex morphology given by $\psi_0=\phi_0=\pi/4$.}
\label{aninon}
\end{figure}

\subsection{Elliptical stigmatic beams}

The anisotropy for an elliptical stigmatic beam is given by
\begin{equation}
\cos\psi(t) = {2 \left[ \cos\psi_0 \gamma^2 t^2 - \sin\psi_0 \sin\phi_0 \gamma (\gamma^2-1) t + \cos\psi_0 \gamma^2 \right] \over \left[ \sin\psi_0 \cos\phi_0 (\gamma^4-1) + \gamma^4 + 1 \right] t^2 + 2 \gamma^2} .
\label{anin}
\end{equation}
If the vortex is canonical in the central plane ($\psi_0=0,\pi$), the anisotropy becomes
\begin{equation}
\cos\psi(t) = {\pm 2\gamma^2 \left( t^2 + 1 \right) \over \left( \gamma^4 + 1 \right) t^2 + 2 \gamma^2} .
\label{anin0}
\end{equation}
Apart from the sign in front of it, the right-hand expression in Eq.~(\ref{anin0}) is positive, which means that there are no flip planes. It therefore requires a non-canonical vortex in the central plane to produce flip planes. To find the minimum required morphology, one can consider where the numerator of Eq.~(\ref{anin}) becomes zero and find the same requirement given in Eq.~(\ref{fpreq}).

In Fig.~\ref{aninon} we show the evolution of the anisotropy for a beam with $\tau=0$ and $\gamma=4$ and a vortex with morphology given by $\psi_0=\phi_0=\pi/4$. One can see that there is a range of propagation distances on one side of the central plane for which the anisotropy is negative, hence, where the topological charge of the vortex is inverted. We also see that the maximum anisotropy is achieved in a plane different from the central plane.

\subsection{Symmetric astigmatic beams}

The anisotropy for a symmetric astigmatic beam is given by
\begin{equation}
\cos\psi(t) = {2\tau\sin\psi_0 \sin\phi_0 t^2 + \cos\psi_0 \left[ (1-\tau^2) t^2+(1+\tau^2)^2 \right] \over (1+\tau^2+t^2+2\tau\sin\psi_0\cos\phi_0 t) (1+\tau^2)} .
\label{anins}
\end{equation}

For $\psi_0\neq 0$ the vortex is not canonical in the central plane. However, it becomes canonical at another point along the propagation direction if
\begin{equation}
\tau = -{\cos\psi_0\pm 1\over\sin\psi_0\sin\psi_0} ,
\end{equation}
where the sign corresponds to the sign of the topological charge of the vortex in the central plane. The location where the vortex becomes canonical is
\begin{equation}
t_+ = -\tan\phi_0-{1\mp\cos\psi_0\over(1\pm\cos\psi_0)\cos\phi_0\sin\phi_0} ,
\end{equation}
where the upper (lower) sign represent a canonical vortex with a positive (negative) topological charge.

\begin{figure}[ht]
\centerline{\scalebox{0.5}{\includegraphics{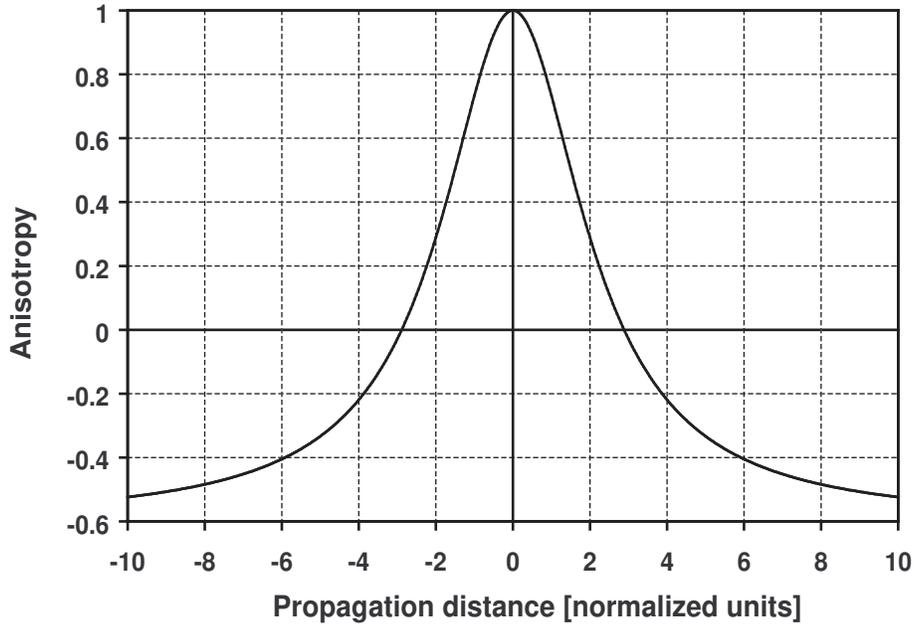}}}
\caption{Anisotropy of a canonical vortex in a symmetric astigmatic beam with $\tau=2$, as a function of propagation distance $t$.}
\label{ansi}
\end{figure}

The anisotropy is shown in Fig.~\ref{ansi} as a function of $t$ for the canonical case with $\tau=2$ and $\gamma=1$. The function is symmetric around the $y$-axis because it only depends on $t^2$. At the central plane the anisotropy is 1, which is the value with which it has been launched. This value decreases and reaches zero at the flip planes. For $t\rightarrow\infty$ the anisotropy approaches the value $-3/5$.

\section{CONCLUSIONS}
\label{concl}

We investigated the properties of the planes in astigmatic and/or elliptical beams where the topological charge of optical vortices are inverted. On the one hand, these flip planes are obtained as the zeros of the denominator of the trajectory equations. On the other hand, they are given as the zeros in the anisotropy of the vortex as a function of the propagation distance. Two special cases are considered: elliptical stigmatic beams and  symmetric astigmatic beams. For the first case flip planes only exist if the vortex is sufficiently anisotropic in the waist (central plane) of the beam. In the second case flip planes exists even for vortices that are canonical in the central plane, provided that the beam is sufficiently astigmatic.

The behavior of vortices and in particular the location of the flip planes in astigmatic and/or elliptical beams can thus be predicted. This provides important information for the design of optical systems in the application of optical vortices.

\end{document}